\documentclass[twocolumn,showpacs,prl,amsmath,amssymb]{revtex4}

\usepackage{graphicx}
\usepackage{dcolumn}
\usepackage{bm}

\begin{document}

\title{ Spatial weak-light solitons in an electro-magnetically 
induced nonlinear waveguide}
\author{Tao Hong}
\address{NTT Basic Research Laboratories, NTT Corporation,\\
3-1, Morinosato-Wakamiya, Atsugi-shi, Kanagawa 243-0198, Japan}

\begin{abstract}
We show that a weak probe light beam can form spatial solitons 
in an electro-magnetically induced transparency (EIT) medium composed 
of four-level atoms and a coupling light field. We find that the coupling 
light beam can induce a highly controllable nonlinear waveguide and exert 
very strong effects on the dynamical behaviors of the solitons. Hence, 
in the EIT medium, it is not only possible to produce spatial solitons 
at very low light intensities but also simultaneously control these 
solitons by using the coupling-light-induced nonlinear waveguide.
\end{abstract}

\maketitle 

Finding new physical systems for producing optical solitons at very 
low light intensities with good controllabilities is very important 
for investigating the nonlinear dynamics of quantum solitons and 
inventing new optical devices for quantum or conventional optical 
communications and computations 
\cite{YamamotoNature,KavisharReport,QiaoQuantumWire}.

All previous research on optical spatial solitons is 
limited to conventional nonlinear mediums 
\cite{KavisharReport,SpatialSolitonNaRb}. In conventional 
nonlinear mediums, due to the Kramers-Kronig relation 
between the refractive index and the absorption coefficient, 
a large nonlinear refractive index is always associated with 
a large absorption, which forms a basic limitation to realizing 
solitons at very low light intensities and studying quantum 
solitons with only a few photons per cross section
\cite{YamamotoNature,KavisharReport,QiaoQuantumWire,SpatialSolitonNaRb}.  
As a result, although spatial solitons have been found to be 
very favorable for applications in optical communications and 
computations \cite{KavisharReport}, it is very hard to extend 
proposals based on spatial solitons to the quantum level to 
tackle problems in quantum communications and computations.  

In comparison with conventional nonlinear mediums, EIT mediums 
have many unique features that are advantageous for producing 
spatial solitons at low light intensities with good 
controllabilities. Recently, EIT mediums have been found not 
only to be able to transmit a light field at ultra-slow light 
speed \cite{HarrisPhysicsToday}, but also to provide very 
large optical nonlinearity to form very strong interactions 
between photons 
\cite{HarrisNonlinearity,ImmamogluOptLett,XiaoPapers,ImmamogluPRL,LukinScience,HongOptComm}.  
In addition to these interesting properties, EIT mediums 
exhibit the following other important features: First, 
the magnitude of the nonlinear coefficient is approximately 
inversely proportional to the coupling light intensity and 
highly controllable in a very large range 
\cite{XiaoPapers,HongOptComm}. Second, both the linear 
absorption and linear refractive index are approximately 
independently controllable, as well as the nonlinear 
susceptibility \cite{XiaoPapers,HongOptComm,FocusPRL}. 
Third, the influence of the EIT mediums on the coupling 
light beam is negligible, which permits us to shape the 
coupling beam with fairly large freedom. In contrast, no 
conventional nonlinear medium has been found to possess 
these properties. 

Here, we will prove that spatial solitons can be formed in
an EIT medium composed of four-level atoms at very low light 
intensities,  analyze some basic limitations on the formation 
of solitons in the medium due to the EIT conditions, and 
predict the experimental possibilities. We will make use 
of the above unique features of EIT mediums to show the 
good controllabilities of the spatial solitons, i.e., very 
strong effects of the coupling-light-induced nonlinear 
waveguide on the spatial solitons. Thus, we provide a new tool, 
i.e., weak-light spatial solitons in the EIT mediums, 
to tackle certain problems of quantum communications 
and computations.
\begin{figure}[h]
\includegraphics[width=3.5cm, height=6cm, angle=270]{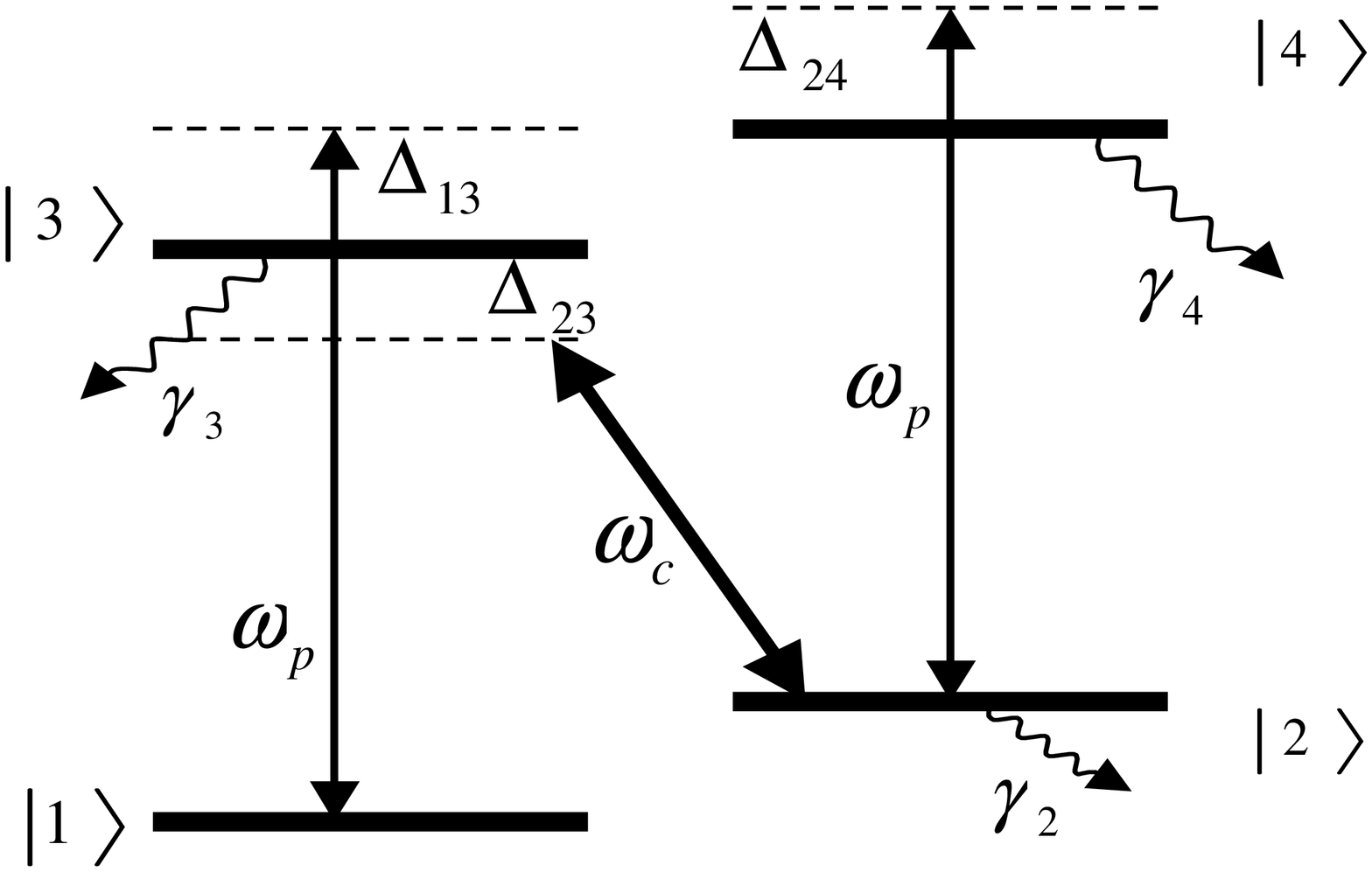} 
\caption{Energy levels and optical couplings of the 
four-level atomic system. $\omega_{c}$ and $\omega_{p}$ 
are the angular frequencies of the coupling light and the 
probe light, respectively. Direct electric-dipole transition 
between two ground states, $|1\rangle$ and 
$|2\rangle$, is forbidden. $\gamma _{2}$, $\gamma_{3}$, 
$\gamma_{4}$ are decay rates of the atomic states. 
$\Delta_{23}$, $\Delta _{13}$ and $\Delta _{24}$ are 
light detunings.} 
\label{fig1}
\end{figure}

Let us consider the propagation of a very weak probe 
light field of frequency $\omega_{p}$ in an EIT medium 
composed of four-level atoms and a strong coupling light 
field of frequency $\omega_{c}$, as shown in Fig. \ref{fig1}.  
According to previous studies of this system, we know that 
when most atoms are in the ground state $\left| 1\right\rangle $, 
the coupling light field can not only induce transparency for 
the probe light field, but also enhance the optical Kerr 
nonlinearity very effectively 
\cite{ImmamogluOptLett,ImmamogluPRL,HongOptComm}. 
From the Maxwell-Bloch equations of the light fields 
and the four-level atomic system, assuming the amplitude 
of the probe light field varying slowly in the $z$-direction, 
we can obtain the following propagation equation for the probe
light field:
\begin{equation}
2ik_{p}\frac{\partial E_{p}}{\partial z}+\nabla _{\bot
}^{2}E_{p}=-k_{p}^{2}\chi E_{p}  \label{eq1}
\end{equation}
where $E_{p}$ and $k_{p}$ are the amplitude (The 
time-dependent traveling wave part is eliminated) and 
the wave vector length of the probe light field, respectively. 
$\chi$ is the susceptibility of the EIT medium, the form of which is
\begin{widetext}
\begin{equation}
\chi (E_{p},E_{c})=-\frac{n\left| \mu
_{13}\right| ^{2}}{\varepsilon _{0}\hbar \Gamma _{3}}\left[ \left( 1-\frac{%
\frac{\left| \Omega _{c23}\right| ^{2}}{\Gamma _{3}}+\frac{\left| \Omega
_{c23}\right| ^{2}\left| \Omega _{p13}\right| ^{2}}{\Gamma _{3}^{2}M}}{M+%
\frac{\left| \Omega _{c23}\right| ^{2}\left| \Omega _{p13}\right| ^{2}}{%
\Gamma _{3}^{2}M}}\right)   
 +\frac{\left| \mu _{24}\right| ^{2}\left| \Omega
_{c23}\right| ^{2}\left| \Omega _{p13}\right| ^{2}}{\left| \mu _{13}\right|
^{2}\Gamma _{4}\Gamma _{3}^{\ast }}\left| \frac{1+\frac{\left| \Omega
_{p13}\right| ^{2}}{\Gamma _{3}M}}{M+\frac{\left| \Omega _{c23}\right|
^{2}\left| \Omega _{p13}\right| ^{2}}{\Gamma _{3}^{2}M}}\right| ^{2}\right] 
\label{eq2}
\end{equation}
\end{widetext}
where $n$ is the atomic density, $\mu_{13}$, $\mu_{24}$ and 
$\mu_{23}$ are electric dipole moments, $E_{c}$ is the 
amplitude of the coupling light field, 
$\left| \Omega _{p13}\right| ^{2}=\left| \mu _{13}\right|^{2}
\left| E_{p}\right| ^{2}/(4\hbar ^{2})$ and $\left| 
\Omega _{p24}\right| ^{2}=\left| \mu _{24}\right|^{2}\left| E_{p}
\right| ^{2}/(4\hbar ^{2})$ are the squared Rabi frequencies 
induced by the probe light and $\left| \Omega _{c23}\right|^{2}=
\left| \mu _{23}\right|^{2}\left| E_{c}\right| ^{2}/(4\hbar ^{2})$ 
is the squared Rabi frequency induced by the coupling light. 
Additionally, $M=\Gamma_{2}+|\Omega_{p24}|^{2}/\Gamma_{4}+
|\Omega_{c23}|^{2}/\Gamma_{3}$, $\Gamma _{3}=\Delta _{13}+
i\gamma _{3}$, $\Gamma _{4}=\Delta _{24}+\Delta_{13}-\Delta _{23}
+i\gamma _{4}$ and $\Gamma _{2}=\Delta _{23}-\Delta_{13}-i\gamma _{2}$. 
In the derivation process \cite{HongSeparatePaper}, we used the 
rotating wave approximation and the adiabatic approximation by 
assuming the EIT condition 
\begin{equation}
\gamma _{3},\gamma _{4},\left| \Omega _{c23}\right| >>\left| \Omega
_{p13}\right| ,\left| \Omega _{p24}\right|   \label{eq3}
\end{equation}
If the conditions 
\begin{equation}
\left|\Gamma _{2}\right|<<\frac{\left| \Omega _{p24}\right| ^{2}}{\left| \Gamma
_{4}\right| }<<\frac{\left| \Omega _{c23}\right| ^{2}}{\left| \Gamma
_{3}\right| }  \label{eq4}
\end{equation}
and 
\begin{equation}
\left| \Delta _{24}+\Delta _{13}-\Delta _{23}\right| >>\gamma
_{4}   \label{eq5}
\end{equation}
are also satisfied, then we can neglect the single-photon and 
two-photon absorption effects. Thus, we simplify Eq.(\ref{eq1}) 
to the well-known nonlinear Schr{\"o}dinger equation (NLSE)
\begin{equation}
2ik_{p}\frac{\partial E_{p}}{\partial z}+\nabla _{\bot
}^{2}E_{p}=C_{n}\left| E_{p}\right| ^{2}E_{p}  \eqnum{(5)}  
\label{eq6}
\end{equation}
where the nonlinear coefficient
$C_{n}=\frac{2\mu _{0}n\left| \mu _{13}\right| ^{2}\left| \mu _{24}\right|
^{2}\omega _{p}^{2}}{\left| \mu _{23}\right| ^{2}\left| E_{c}\right| ^{2}
\mathop{\rm Re}
(\Gamma _{4})\hbar }$
This is a (2+1)-dimensional NLSE, which has many classes of
nonlinear solutions that describe various self-sustained structures, such as
self-focused light beams, optical vortices and quasi-(1+1)-dimensional 
optical bright and dark solitons etc. 
\cite{KavisharReport,SolitonBooks,ZakharovJETP}.  Because these 
solutions have many common features, in this paper we only consider 
one typical class, i.e., the quansi-(1+1)-dimensional bright solitons
 in the case of ${\rm Re}(\Gamma_{4})<0$. The form of a fundamental 
bright soliton is
\begin{equation}
E_{p}=2m E_{0}{\rm sech}(\sqrt{2\left|C_{n}\right|}
mE_{0}x)\exp (\frac{im^{2}E_{0}^{2}
\left|C_{n}\right|}{k_{p}}z) 
\label{eq7}
\end{equation}
where $m$ is the eigen value of the soliton, 
$E_{0}=\hbar \gamma/\mu$ (for simplicity, we let 
$\mu _{13}=\mu _{24}=\mu _{23}=\mu $ and 
$\gamma _{4}=\gamma_{3}=\gamma $). The maximum amplitude of 
the soliton is 
$E_{p\max }=2m\gamma \hbar /\mu   \eqnum{(8)}$, and it is 
related to the width of the soliton, 
\begin{equation}
x_{_{\rm FWHM}}=2{\rm ln}(2+\sqrt{3})\sqrt{
\frac{\varepsilon _{0}\left|{\rm Re}(\Gamma _{4})\right|\hbar}
{n\mu^{2}k_{p}^{2}}}\frac{E_{c}}{E_{p\max}}
\label{eq8}
\end{equation}
Thus, we find that the width of the soliton is actually determined by
the amplitude ratio between the coupling light and the probe light 
instead of by the amplitude of each light. This indicates that as 
long as Eqs.(\ref{eq3})-(\ref{eq4}) can be well satisfied, we can
 reduce the coupling light amplitude gradually along the propagation
 direction to keep the soliton width constant even if there remains 
some small absorption in the medium that attenuates the probe light
 intensity and hence leads to broadening of the soliton width 
\cite{SpatialSolitonNaRb,TwoPhotonAbsorption}. Apparently, in a 
conventional nonlinear medium, we have no choice like this but to 
introduce a gain process \cite{SolitonBooks}. Additionally, it 
indicates that as long as the decay rate $\gamma_{2}$ is small enough, 
it is possible to realize these spatial solitons with finite width at 
very low probe light intensities, which is exactly what we mean by 
spatial weak-light solitons in this paper.

Because the NLSE is obtained under the conditions defined by 
Eqs.(\ref{eq3})-(\ref{eq5}),
the soliton solution should also satisfy them. Thus, these
requirements set fundamental limitations on the amplitude and 
the width of the soliton such that,
\begin{equation}
E_{p\max }=\frac{2m\gamma \hbar }{\mu }<<\min (\frac{\gamma \hbar }{\mu }%
,E_{c})  \label{eq9}
\end{equation}
and
\begin{equation}
x_{_{\rm FWHM}}>>\frac{2{\rm ln}(2+\sqrt{3})}{\sqrt{\frac{n\mu ^{2}}{\varepsilon
_{0}\gamma \hbar }}k_{p}}  \label{eq10}
\end{equation}
Eq. (\ref{eq9}) indicates that the soliton can only be realized at  
weak light intensities, and Eq.(\ref{eq10}) indicates that it is 
impossible to produce a soliton with an arbitrary narrow width. 
Before further discussion, we give a rough numerical estimate of 
the width and photon flux of the soliton. We assume an atomic medium 
of density $n=1\times 10^{14}cm^{-3}$, typical dipole matrix elements 
for alkali atoms $\mu =3\times 10^{-29}C\cdot m$, the decay rates of 
upper excited states $\gamma =30MHz$, the maximum amplitude of the 
probe light $E_{p\max}=10^{-2}E_{c}$, wavelength of the probe light 
$\lambda _{p}=800nm$ and ${\rm Re}(\Gamma _{4}) =-10\gamma $, 
and then we can obtain $x_{_{\rm FWHM}}\thickapprox0.06 mm$. 
If $E_{c}=\gamma\hbar/\mu$, then the photon flux of the probe light is
about $6 mm^{-2}ns^{-1}$. This rough estimation shows that a soliton 
can be produced with observable quantum properties within current 
experimental achievable range. 

\begin{figure}[h]
\includegraphics[width=6cm, height=10cm, angle=270]{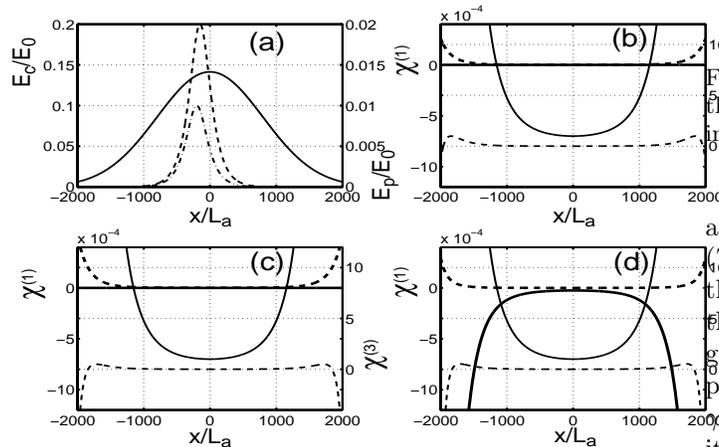}
\caption{(a) shows the transverse profile of the coupling light beam (solid
curve) and the fundamental soliton (dot-dashed curve) or second-order
soliton formed by the probe light beam. (b), (c) and (d) show the transverse
structures of the coupling-light-induced nonlinear waveguide: linear 
susceptibility $\chi ^{(1)}$ (thick lines) and
third-order susceptibilities $\chi ^{(3)}$ (thin lines), in which solid
lines present the real parts and the dashed lines present the imaginary
parts, under different conditions, respectively. $\chi ^{(1)}$ is in units
of $\mu^{2}n/(\varepsilon_{0}\hbar \gamma )$. $\chi ^{(3)}$ is in units of 
$\mu^{2}n/(\varepsilon_{0}\hbar \gamma E_{0}^{2})$. } 
\label{fig2}
\end{figure}

The unique features of an EIT medium enable the coupling light 
beam to induce a very special nonlinear waveguide, which can 
exert very strong effects on the dynamical behaviors of
solitons formed by the probe light beam and hence give rise to new
soliton controllabilities. Next, solving Eq.(\ref{eq1}) (neglecting the
coordinate y) numerically by using the Crank-Nicolson method, we show
several typical effects of the nonlinear waveguide on the evolution of
(1+1)D solitons in the EIT medium. In the calculation, we use $E_{0}=\gamma\hbar/\mu$,
$L_{a}=\sqrt{\varepsilon_{0}\gamma\hbar/(n\mu^{2}k_{p}^{2})}$ and 
$L_{b}=2\varepsilon_{0}\gamma\hbar/(n\mu^{2}k_{p})$ as units to normalize 
the amplitudes of the
electro-magnetic fields and the spatial coordinates $x$ and $z$, 
respectively.

\begin{figure}[h]
\includegraphics[width=4cm, height=8cm, angle=270]{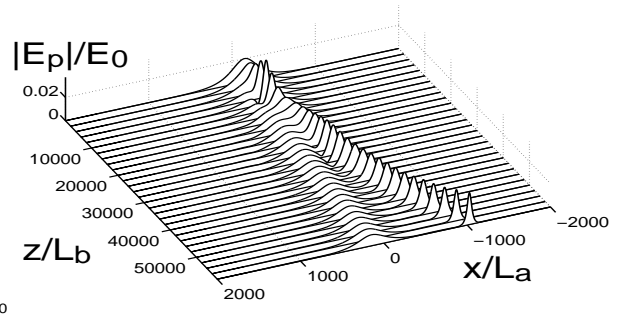}
\caption{Second-order soliton splits into two solitons under the influence of
the transverse variation of the coupling-light-induced nonlinear
susceptibility [see Fig. \ref{fig2} (b)].} 
\label{fig3}
\end{figure}

Because the coupling light intensity is approximately inversely proportional
to the nonlinear susceptibility under the EIT condition, the
coupling-light-induced transverse variation of the nonlinear susceptibility
can strongly affect the dynamical behaviors of high-order solitons in the
EIT medium. For example, the gradient of the real part of the
coupling-light-induced third-order susceptibility, as shown by the thin
solid curve in Fig. \ref{fig2} (b), can lead to splitting of a second-order soliton
[The maximum amplitude is twice that of the soliton described by Eq. (\ref{eq7})],
 as shown in
Fig.\ref{fig3}. In this calculation, we assume that the coupling light beam has a
Gaussian profile along the $x$ axis [See the solid curve in Fig.\ref{fig2}(a)], 
and propagates along the z axis with negligible diffraction. Other parameters
are as follows: $\Delta _{13}=\Delta _{23}=0\gamma $, $\Delta
_{23}=100\gamma $, $\gamma _{2}=1\times 10^{-8}\gamma $. The linear 
and nonlinear susceptibilities induced by this
light beam are shown in Fig.\ref{fig2}(b). The initial second-order soliton formed
by the probe light beam has a small transverse displacement from the center
of the coupling beam, as shown by the dashed line in Fig.\ref{fig2}(a), hence it
experiences a non-uniform nonlinear phase modulation due to the gradient of
the third-order susceptibility. As we known, if a second-order soliton
propagates in a uniform Kerr nonlinear medium, it only exhibits periodic
oscillation \cite{SolitonBooks}. The splitting of the second-order
soliton into two solitons is therefore just due to the transverse variation
of the nonlinear susceptibility. Because the split solitons have opposite
transverse momentums, one split soliton propagates into the lower coupling
intensity regime, i.e., the larger nonlinearity region; it becomes
narrower. The other one propagates to the higher coupling intensity regime,
i.e., the smaller nonlinearity region; it becomes wider. 

\begin{figure}[h]
\includegraphics[width=4cm, height=8cm, angle=270]{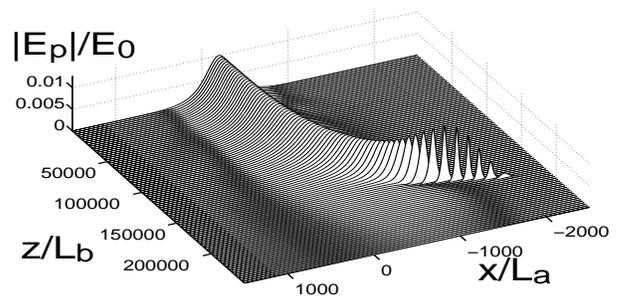}
\caption{Absorption of the soliton due to the increasing linear absorption at
the boundary of the coupling beam.} 
\label{fig4}
\end{figure}

At the boundary of the coupling beam, the linear absorption due to
ground-state dephasing increases very fast. As a result, if a soliton
propagates toward the boundary of the coupling beam, its amplitude will
decay very quickly because of the absorption, as shown in Fig. \ref{fig4}. In this
calculation, we slightly increase the decay rate of the gound state $\left|
2\right\rangle $, i.e.,  $\gamma _{2}=2\times 10^{-8}\gamma $, to make the
linear absorption more apparent, as indicated by the imaginary part of the
linear susceptibility (thick dashed line) in Fig. \ref{fig2}(c). The initial fundamental
soliton with zero transverse velocity has a larger offset from the center
of the coupling beam, as plotted by the dot-dashed line in Fig. \ref{fig2}(a). Due to
the gradient of the nonlinear susceptibility, the soliton is transversely
accelerated and propagates toward the boundary of the coupling beam. Because
of the linear absorption becomes larger and larger, the amplitude of the
soliton is quickly reduced. However, the width of the soliton becomes
smaller in comparison with its initial value, although the amplitude becomes
smaller because the nonlinear susceptibility becomes larger relatively
more rapidly in this specific condition. 

\begin{figure}[h]
\includegraphics[width=4cm, height=8cm, angle=270]{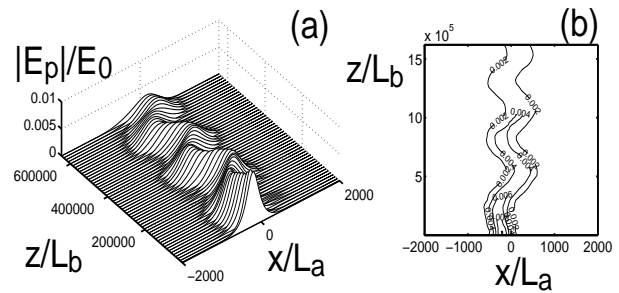}
\caption{Nonlinear oscillation of the soliton due to the transverse
confinement produced by the coupling-light-induced non-uniform linear
and nonlinear susceptibilities [See Fig. 2 (d)]. Here (b) is
the contour graph of (a).} 
\label{fig5}
\end{figure}

If the detuning $\Delta _{13}$ is finite, the coupling beam can form a
high-linear-refractive-index waveguide overlapping with the nonlinear one
and force the soliton to oscillate nonlinearly inside the waveguide \cite{FocusPRL}. For
example, we follow the above parameters values except for setting $\Delta
_{13}=-5\times 10^{-7}\gamma $, $\gamma _{2}=5\times 10^{-9}\gamma $,
and we find that the real part of the linear susceptibility is much reduced 
at the boundary of the coupling beam [See the thick solid line in 
Fig. \ref{fig2}(d)]. Thus, the linear susceptibility behaves as a potential, 
which confines the soliton to oscillate, as shown in Fig. \ref{fig5}. The
oscillation is nonlinear because of the nonuniform nonlinear susceptibility.
For example, the initial transversely static soliton is accelerated by the
gradient of the nonlinear susceptibility because the amplitude of the
soliton is very large and the soliton propagates toward the boundary of the
coupling beam, as shown in Fig. \ref{fig5}(b). But, because the linear 
absorption becomes larger and larger, the intensity of the soliton become 
smaller and smaller. As a result, the acceleration due to the nonlinear 
susceptibility becomes
smaller and smaller. When the acceleration due to the linear
susceptibility overcomes the nonlinear one, the soliton is gradually 
decelerated, and finally reverses its propagation direction and oscillates 
around the center of the coupling beam. 

In conclusion, we have shown a weak probe light beam can form spatial 
solitons in an electro-magnetically induced transparency medium composed 
of four-level atoms and a coupling light field. We have found that 
the coupling light beam can induce a highly controllable nonlinear 
waveguide and exert very strong effects on the dynamical behaviors of
 the solitons. Hence, in the EIT medium, it is not only possible to 
produce spatial solitons at very low light intensities but also 
simultaneously control these solitons by using the coupling-light-induced 
nonlinear waveguide.

I thank Makoto Yamashita and Michael Wong Jack for their very helpful discussions.

\end{document}